\documentclass[aps,superscriptaddress,reprint,twocolumn]{revtex4-1}
\usepackage[colorlinks, citecolor=blue, linkcolor=blue]{hyperref}
\usepackage[english]{babel}
\usepackage{amsmath}
\usepackage{amssymb}
\usepackage{mathtools}
\usepackage{array}
\usepackage{enumitem}
\usepackage{nicefrac}
\usepackage[separate-uncertainty = true]{siunitx}
\usepackage{graphicx}
\usepackage{xcolor}
\usepackage{xspace}
\usepackage{changes}
\newlist{todolist}{itemize}{2}
\setlist[todolist]{label=$\square$}
\usepackage{pifont}
\newcommand{\BFO}{BiFeO$_3$\xspace}

\begin{document}

\title{Engineering magnetic domain wall energies in multiferroic BiFeO\textsubscript{3} via epitaxial strain}

\author{Sebastian Meyer}
\thanks{These two authors contributed equally}
\email[E-mail: ]{smeyer@uliege.be}
\affiliation{Nanomat/Q-mat/CESAM, Universit\'e de Li\`ege, B-4000 Sart Tilman, Belgium}
\affiliation{TOM/Q-mat/CESAM, Universit\'e de Li\`ege, B-4000 Sart Tilman, Belgium}
\affiliation{Fonds de la Recherche Scientifique (FRS-FNRS), Bruxelles, Belgium}

\author{Bin Xu}
\thanks{These two authors contributed equally}
\affiliation{Jiangsu Key Laboratory of Thin Films, School of Physical Science and Technology, Soochow University, Suzhou 215006, China}
\affiliation{Physics Department and Institute for Nanoscience and Engineering, University of Arkansas, Fayetteville, Arkansas 72701, USA}


\author{Laurent Bellaiche}
\affiliation{Physics Department and Institute for Nanoscience and Engineering, University of Arkansas, Fayetteville, Arkansas 72701, USA}

\author{Bertrand Dup\'e}
\affiliation{Nanomat/Q-mat/CESAM, Universit\'e de Li\`ege, B-4000 Sart Tilman, Belgium}
\affiliation{TOM/Q-mat/CESAM, Universit\'e de Li\`ege, B-4000 Sart Tilman, Belgium}
\affiliation{Fonds de la Recherche Scientifique (FRS-FNRS), Bruxelles, Belgium}

\date{\today}

\begin{abstract}
Epitaxial strain has emerged as a powerful tool to tune magnetic and ferroelectric properties in functional materials such as in multiferroic perovskite oxides. 
Here, we use first-principles calculations to explore the evolution of magnetic interactions in the antiferromagnetic multiferroic BiFeO\textsubscript{3} (BFO), one of the most promising multiferroics for future technology. The epitaxial strain in BFO(001) oriented film is varied between $\varepsilon_{xx,yy} \in [-2\,\%, +2\,\%]$. We find that both strengths of the exchange interaction and Dzyaloshinskii-Moriya interaction (DMI) decrease linearly from compressive to tensile strain whereas the uniaxial magnetocrystalline anisotropy follows a parabolic behavior which lifts the energy degeneracy of the (111) easy plane of bulk BFO.
From the trends of the magnetic interactions we can explain the destruction of cycloidal order in compressive strain as observed in experiments due to the increasing anisotropy energy. For tensile strain, we predict that the ground state remains unchanged as a function of strain.
By using the domain wall (DW) energy, we envision the region where isolated chiral magnetic texture might occur as function of strain i.e. where the DW and the spin spiral energy are equal. This transition between $-1.5\,\%$ and $-0.5\,\%$ of strain should allow topologically stable magnetic states such as antiferromagnetic skyrmions and merons to occur. Hence, our work should trigger experimental and theoretical investigations in this range of strain.
\end{abstract}

\maketitle

\section{Introduction}

For over a decade, topologically stable magnetic structures, such as domain walls (DWs) and skyrmions have been studied extensively based on their theoretical prediction in the early 90's \cite{Bogdanov1989,Bogdanov1994} and their first experimental observation 20 years later \cite{Muhlbauer2009,Yu2010}. Their potential in low-power, high-density information storage, spintronic applications and their unique stability, resistance to external magnetic fields, and efficient dynamics make them a fascinating area of study \cite{Kiselev2011,Fert2013,Parkin2015,Back2020}. The formation and control of these skyrmions have primarily been explored in ferromagnetic materials, but their antiferromagnetic counterparts are predicted to exceed their potential \cite{Jungwirth2016,Zhang2016,Barker2016,Jungwirth2018}. While the stabilization of skyrmions in FMs can be achieved e.g. by the application of an external magnetic fields \cite{Romming2013,Moreau-Luchaire2016,Soumyanarayanan2017,Boulle2016,Herve2018,Pollard2017}, antiferromagnets are unaffected by those. Therefore, the stabilization of chiral magnetic textures in antiferromagnets remains a challenge and depends on the precise control of magnetic interactions.

Epitaxial strain allows for precise modification of a material's lattice structure through heteroepitaxial growth and has emerged as a powerful tool to engineering magnetic properties in materials. For example, it offers the means to tailor crystal symmetries \cite{Hatt2010-tk,Dupe2011-of}, magnetic anisotropies, and exchange interactions and even to control the magnetic ground state of a material \cite{Sando2013}. In this way, skyrmions could be stabilized in ferromagnetic (FM) systems \cite{Hu2018,Wang2018,Budhathoki2020,Zhang2021}. Such a technique can pave the way to the stabilization and control of magnetic structures such as skyrmions in antiferromagnets. Strain engineering has already been used in multiferroic oxides, where several ferroic orders, such as the polarization and the magnetization, can be tuned \cite{Govinden2023,Qiang2020,Hemme2023,Dufour2023}. More interestingly, their coupling called the magneto-electric coupling is affected by strain. Since they do not only posses a magneto-elastic, but also a magneto-electric and piezo-electric coupling, the magnetism can be tuned in various ways. To that end, one of the most promising multiferroic material is BiFeO\textsubscript{3} (BFO) where ferroelectricity and antiferromagnetism can coexist far beyond room temperature \cite{Roginskaya1966,Neaton2005,Ravindran2006,Kornev2007,Haumont2008}. 

Besides being extensively studied in the bulk, the interest in strained BFO has also risen up due to the variety of non-collinear states that it can host. Bulk BFO has an antiferromagnetic (AFM) spin spiral ground state - the so-called type-I cycloid of 62~nm pitch \cite{Sosnowska1982,Sosnowska1996}. This spin spiral can be destroyed via biaxial strain when BFO thin film is grown on various substrates \cite{Bai2005,Bea2007,Zhao2006,Ratcliff2011,Sando2013} or by uniaxial strain \cite{Hemme2023} which suggests a high magneto-elastic response. This response can affect all magnetic interactions in BFO e.g. the magnetic exchange, the anisotropy and the Dzyaloshinskii-Moriya interaction (DMI) whom ratios have yet to be quantified via density functional theory (DFT). A region of particular importance is the transition between collinear and non-collinear magnetic ground states as function of strain because more complex isolated magnetic textures, such as skyrmions, antiskyrmions or merons/anti-merons can form.

In this paper, we explore the different magnetic interactions of BiFeO\textsubscript{3} under the influence of bi-axial strain based on DFT. Varying the nominal strain $\varepsilon_{xx,yy} \in [-2\,\%,+2\,\%]$ from compression to elongation, we determine the magnetic exchange interaction, Dzyaloshinskii-Moriya interaction and the anisotropy energy. We study the influence of epitaxial strain on the magnetism in the $Cc$ phase of BFO to identify the region of strain, where the transition from non-collinear to collinear order takes place. We find that both exchange and DMI decrease linearly with strain whereas the anisotropy follows a parabolic behavior, preferring the $[11\overline{2}]$ direction for compressive strain and the $[1\overline{1}0]$ direction for tensile strain.
Following the ratios of magnetic interactions, we conclude that with compressive strain, collinear G-type antiferromagnetic order is enhanced due to an increase of the anisotropy and magnetic exchange over the DMI. In the tensile regime, the results are ambiguous. From our calculations, we expect that the ground state spin spiral of the bulk $R3c$ phase in BFO is preserved; however, since the energies are extremely close, small distortions or impurities can also favor collinear order.
At last we analyze the evolution of the domain wall energy to locate the region of strain where chiral non-collinear magnetic textures such as antiferromagnetic skyrmions and merons could be stabilized in BiFeO\textsubscript{3}. We identify the transition between a non-collinear and a collinear ground state in BFO in between $-1.5\,\%$ and $-0.5\,\%$ of compressive strain where such states should occur. The presented work should therefore trigger experimental and theoretical investigations in that range of strain.

This paper is structured as follows: In Sec.~\ref{Section: Methods}, we describe our procedure for the DFT calculations. In Sec.~\ref{Section: Results}, the results of the DFT calculations are presented which are discussed in Sec.~\ref{Section: Discussion}. A conclusive statement is given in Sec.~\ref{Section: Conclusion}.

\section{Methods} \label{Section: Methods}

\subsection{Structural Relaxation}

\begin{figure}
\includegraphics[scale=1]{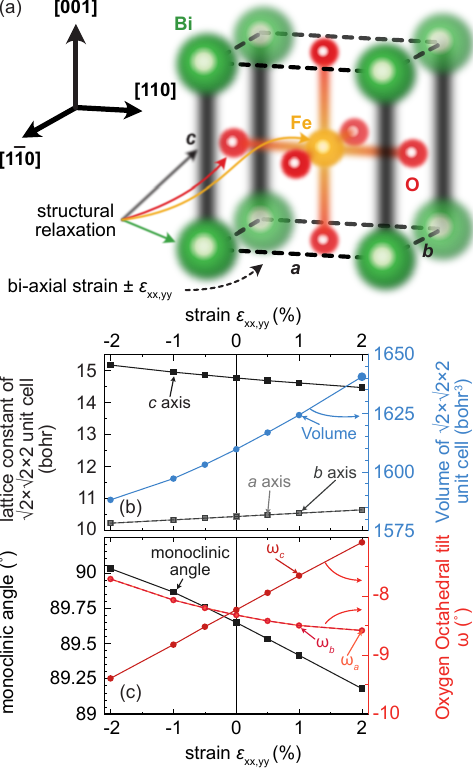}
\caption{Structural properties of strained BiFeO\textsubscript{3}. (a) Sketch of strained $Cc$ BFO in cubic representation: Bi-axial strain $\varepsilon$ is applied in the $xy$ plane with an angle of $90^\circ$ between the $a$ and $b$ axis, whereupon the other parameters ($c$ axis, monoclinic angle and all atomic positions) are relaxed - indicated by blurred colors (for more details, see text). (b) Relaxed lattice constant ($a$, $b$ and $c$ axes, squares on left ordinate) vs. value of strain. Right ordinate: Unit cell volume (blue filled circles). Note that the relaxation has been done in a $\sqrt{2}\times \sqrt{2} \times 2$ unit cell - the values are presented within this notation. (c) Left ordinate: Monoclinic angle of the $c$ axis (black squares) and components of the oxygen octahedral tilts $\omega_i$ around the $i$ axis (red circles for the second ordinate). Note that due to the imposed $Cc$ phase at $0\,\%$ strain, the tilt angles around the three axes are not exactly on top of each other due to shear strain in the $xy$ plane.}
\label{Figure: Strain vs Structure}
\end{figure}

We have relaxed all strained structures of \BFO using the Abinit package \cite{ABINIT,Gonze2020} and the projector augmented wave (PAW) method \cite{Bloechl1994}. Therefore, strain $\varepsilon = \frac{a' - a_\textrm{BFO}}{a_\textrm{BFO}}$ is applied from $-2\%$ to $2\%$ for the in-plane axes of the $Cc$ phase of BFO. Here, $a'$ is the varied in-plane lattice constant and $a_\textrm{BFO}$ denotes the BFO bulk lattice constant of the $R3c$ phase. Note that the $Cc$ phase is imposed at $\varepsilon = 0\,\%$ with the lattice constant of $R3c$ BFO.  While $\varepsilon$ is varied from compressive strain of $-2\,\%$ to tensile strain of $+2\,\%$, the other structural parameters are relaxed [indicated by blurred colors of the atoms and bonds, cf. sketch in Fig.~\ref{Figure: Strain vs Structure} (a)].

 We have used a $\sqrt{2}\times\sqrt{2}\times2$ unit cell hosting 20 atoms. The magnetic configuration is chosen as the collinear G-type antiferromagnetic order. We have converged the forces on each atom down to $1\times 10^{-5}\,\textrm{hartree/bohr}$. The exchange and correlation functional is treated with local spin density approximation $+U$ (LSDA$+U$ ), with a Hubbard $U = 4.0\,\textrm{eV}$ parameter and $J=0.4\,\textrm{eV}$ on the Fe atoms, which are typical values for first-principles calculations on BFO \cite{Kornev2007,Paillard2016,Xu2021}. The wave-functions are expanded with a plane-wave basis set using a kinetic energy cutoff of $E_\textrm{cut} = 30\,\textrm{hartree}$. By using a $k$-point-mesh for the structural relaxation of $8\times 8\times 6$, we obtain a magnetic moment of the Fe atoms of $\vert m_\textrm{Fe}\vert = 4.0\,\mu_B$ for all relaxed structures. The results of the relaxation are presented in Fig.~\ref{Figure: Strain vs Structure} (b,c) and are in accordance with previous studies \cite{Hatt2010-tk,Dupe2011-of}.

\subsection{Determination of Magnetic Interactions}

We use the relaxed structures as described above to determine the magnetic interactions, such as the Heisenberg exchange interaction, the Dzyaloshinskii-Moriya (DM) interaction and the uniaxial anisotropy energy for each value of strain. The first two interactions are determined by calculating energy dispersions of flat homogeneous spin spiral states without (for the exchange) and with (for the DMI) spin-orbit coupling. For both interactions, we apply the pseudo-cubic approximation to compare the results we have obtained for the relaxed $R3c$ structure \cite{Xu2021,Meyer2023}.

Spin spirals are the general solution of the Heisenberg model on a periodic lattice and can be characterized by the spin spiral vector $\mathbf{q}$. This vector determines the propagation direction of the spin spiral as well as the canting angle between two neighboring spins. A magnetic moment $\mathbf{m}_i$ at an atom position $\mathbf{r}_i$ is given by
\begin{equation}
    \mathbf{m}_i = m \Big( \cos (\mathbf{q}\cdot \mathbf{r}_i),\sin (\mathbf{q}\cdot \mathbf{r}_i),0 \Big) \label{Equation: Flat spin spiral m_i}
\end{equation}
where $m$ is the magnitude of the magnetic moment.

The energy dispersions $E(\mathbf{q})$ are obtained applying the full-potential linearized augmented plane wave (FLAPW) approach \cite{krakauer1979,wimmer1981,weinert1982}, as implemented in the \textsc{fleur} code \cite{FLEUR}.
For all these calculations, we have used the LSDA+$U$ \cite{VWN} with a Hubbard $U = 4.0\,\textrm{eV}$ parameter and $J=0.4\,\textrm{eV}$ on the Fe atoms. Muffin-tin radii were set to 2.80 bohrs, 2.29 bohrs, and 1.29 bohrs for Bi, Fe, and O atoms, respectively and a large plane-wave cutoff $k_{\textbf{max}}$ of 4.6 $\textrm{bohr}^{-1}$. These parameters result in a magnetic moment of $m=4.0 \,\mu_{\mathrm{B}}$ for all phases in agreement with experiments \cite{Sosnowska1996}. Calculations along the full paths of the Brillouin zone (BZ) without SOC have been performed self-consistently using the generalized Bloch theorem \cite{Sandratskii1991} and a $k$-point mesh of $10\times 10 \times 10$. To accurately determine the energies around the magnetic ground state (R points of the BZ) at $\vert \mathbf{q} \vert \rightarrow \textrm{R}$, the magnetic force theorem \cite{ForceTheorem,Oswald1985} has been applied using a dense $k$-point set of 64000 $k$ points (i.e., $40\times 40 \times 40$).
The energy dispersion without SOC is interpreted using the Heisenberg exchange interaction.
For the energy contribution due to SOC, $\Delta E_{\textrm{SOC}}$, we add SOC in first-order perturbation theory \cite{Heide2009,Meyer2017} for every previously calculated point. The resulting curve has been interpreted with the Dzyaloshinskii-Moriya interaction \cite{Dzyaloshinskii1958,Moriya1960}.

We determine the Heisenberg exchange interaction constants $J_{ij}$ beyond nearest neighbors by mapping the Heisenberg Hamiltonian 
\begin{equation}
\mathcal{H}_\textrm{ex} = -\sum_{ij} J_{ij} (\mathbf{m}_i \cdot \mathbf{m}_j)
\end{equation}
onto the energy dispersion $E(\mathbf{q})$ of flat spin spiral states neglecting spin-orbit coupling (SOC). We include seven neighbors for the exchange interaction using the pseudo-cubic approximation  for all values of strain. To simplify our model, we approximate the curvature of the combined exchange interactions $J_{1,...,7}$ at $\mathbf{q} \rightarrow R,\, \vert \mathbf{q} \vert \in [0,0.05]$ with an effective nearest neighbor exchange interaction $J_\textrm{eff}$.

For the Dzyaloshinskii-Moriya (DM) interaction, we use the calculated data of the energy contribution due to SOC $\Delta E_\textrm{SOC}$. Therefore, we use the converse spin-current (sc) model,
\begin{equation}
\mathcal{H}_\textrm{SC} = -\sum_{ij} C_{ij} \left( \mathbf{u} \times \mathbf{e}_{ij} \right) \cdot \left( \mathbf{m}_i \times \mathbf{m}_j \right ) \label{Equation: SC model}
\end{equation} 
where $C_{ij}$ determines the strength of the DMI, $\mathbf{u}$ is the unit vector along the polarization direction and $\mathbf{e}_{ij}$ is the unit vector between magnetic sites $i,j$. We include three neighbors for the sc DMI in pseudo-cubic approximation. For the effective DMI, $C_\textrm{eff}$, we linearly map Eq~\eqref{Equation: SC model} to the SOC contribution at $\mathbf{q} \rightarrow R$.

We also have calculated the magnetic anisotropy in \BFO. For that, we have assumed the collinear G-type AFM state for each structure, where we have applied SOC in different directions of the respective BZ. We have used $20\times 20 \times 20$ $k$-point-mesh in the whole Brillouin zone and applied the second quantization with the Force Theorem \cite{ForceTheorem} to determine the energy differences between the hard axis ($[111]$ axis) and other directions as shown in the results. For the description of the domain wall energy, we will use an effective anisotropy $K_\textrm{eff}$ that is the lowest energy of the respective direction with strain.

This method for calculating magnetic interaction in BFO has also been described in detail in Refs.~\onlinecite{Meyer2023,Xu2021}.

\subsection{Error Estimation}

In the following, we will analyze small energy differences that qualitatively affect the ground state of BiFeO\textsubscript{3}. Especially for the DFT calculations, we reach the limit of accuracy below $\sim 0.1\,\textrm{meV}$. It is impossible to estimate the error in DFT calculations itself stemming mostly from the exchange correlation functional simply because the amount of necessary calculations for that is beyond the scope of any DFT study that is not addressed to that specific problem. Hence, the absolute values in our calculations may vary depending on the functional, we can only conclude on the trends. The LSDA+$U$ approximation with the functional of Ref. \onlinecite{VWN} has been used previoulsy in Refs.~\onlinecite{Meyer2023,Xu2021}.

For the effective magnetic interactions including a fitting procedure in the first step, namely the exchange and the DMI, we will address the error based on the error of the fits. For the DMI, the DFT data is perfectly linear and the value $C_\textrm{eff}$ is determined with a negligible error of less than 0.01~\%.
The values for the exchange interactions $J_i$ on the other hand are fitted with an error $\Delta J_i$. This error affects our estimate on $J_\textrm{eff}$ because it describes this energy dispersion curve within $\mathbf{q} \rightarrow R$. Since the energy is the sum for all $J_i$, we apply the sum rule to estimate the error on $J_\textrm{eff}$ as
\begin{equation}
    \Delta J_\textrm{eff} = \frac{1}{6} \sqrt{\sum_{i=1}^7 (\Delta J_i)^2 N_i} \label{Equation: Error Sum rule}
\end{equation}
where $N_i$ are the number of neighbors in the $i$-th shell, which for $J_1$ are 6 neighbors in the cubic lattice. The fraction $\nicefrac{1}{6}$ in front of the square root accounts for the number of the nearest neighbors.

Using multiplication or division, we will use the error estimate according to
\begin{equation}
    \Delta Q = \vert Q \vert \sqrt{\left ( \frac{\Delta J_\textrm{eff}}{J_\textrm{eff}} \right )^2} =\vert Q \vert \frac{\Delta J_\textrm{eff}}{J_\textrm{eff}}  \label{Equation: Error Multiplication}
\end{equation}
where $Q$ is the quantity we determine.

\section{Results} \label{Section: Results}

Figure \ref{Figure: Magnetic trends} shows the trends of magnetic interactions in \BFO (BFO) under the application of bi-axial strain.

\subsection{Exchange Interaction}

\begin{figure*}
\includegraphics[scale=1]{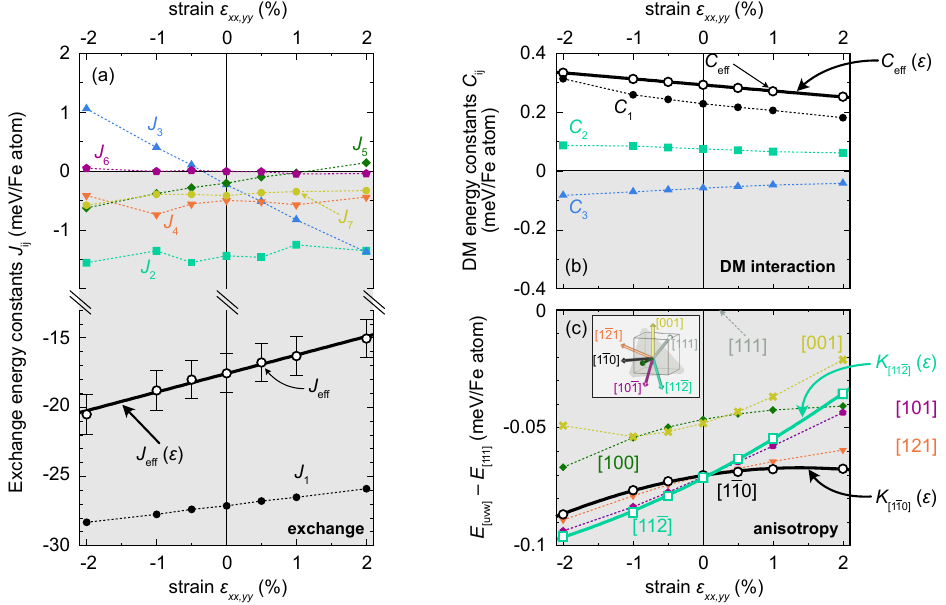}
\caption{Trend of magnetic interactions in \BFO under bi-axial strain. (a) Magnetic Heisenberg exchange interaction parameters $J_{1 \dots 7}$ from pseudo-cubic approximation for different values of strain. Note that the nearest neighbor exchange $J_1$ is dominant and the effective nearest neighbor exchange $J_\textrm{eff}$ matches the curvature of a energy dispersion at $\mathbf{q} \rightarrow R$. The error on $J_\textrm{eff}$ is stemming from $\Delta J_i$ as described in the text. (b) Evaluated spin-current driven Dzyaloshinskii-Moriya interaction ($C_{1\dots 3}$ for the first three nearest neighbors and in effective nearest-neighbor approximation $C_{\textrm{eff}}$), (c) Energy differences for uniaxial anisotropy calculations of different magnetization directions with respect to the [111] direction - the inset shows a sketch with all directions in pseudo-cubic approximation. Note that in the absence of strain, magnetization directions within the (111) plane are energetically degenerate. In all graphs, dashed lines serve as guide to the eye, bold solid lines $J_\textrm{eff}(\varepsilon)$,  $C_\textrm{eff}(\varepsilon)$ and $K_{[uvw]}(\varepsilon)$ are linear fits for $J_\textrm{eff}$, $C_\textrm{eff}$ and quadratic fits for the preferred anisotropy directions respectively.}
\label{Figure: Magnetic trends}
\end{figure*}

For all applied values of strain, the nearest neighbor Heisenberg exchange interaction $J_1$ [black points in Fig.~\ref{Figure: Magnetic trends}~(a)] contains the dominant exchange contribution and prefers collinear antiferromagnetic order. Its strength decreases linearly about $10\,\%$ from $-28\,\textrm {meV/Fe atom}$ at $\varepsilon=-2\,\%$ to $-26\,\textrm{meV/Fe atom}$ at $\varepsilon=+2\,\%$.
The second largest contribution stems from the second neighbor $J_2 \approx -1.5\,\textrm{meV/Fe atom}$ (light green squares). Its strength does not vary drastically with strain, but it leads to exchange frustration with strain as the ratio ${J_2}/{J_1}$ increases.
The largest variation can be seen for the third neighbor exchange interaction $J_3$ [blue triangles in Fig.~\ref{Figure: Magnetic trends}~(a)]. For compressive strain, it prefers ferromagnetic (FM) alignment whereas for tensile strain, it prefers AFM alignment. It linearly decreases from $J_3(\varepsilon = -2\,\%) = 1\,\textrm{meV/Fe atom}$ to $J_3(\varepsilon = +2\,\%) = -1\,\textrm{meV/Fe atom}$.
Neighbors beyond the third shell in the system hold a minor contribution ($\vert J_{4,5,6,7} \vert/ \vert J_1 \vert \leq 0.03$) and do not change significantly with strain. Even though $J_5$ undergoes a sign change it only appears as a slight correction in our method.

To simplify the model, we evaluate the effective nearest neighbor exchange $J_\textrm{eff}$ as shown in black circles of Fig.~\ref{Figure: Magnetic trends}~(a). It reduces its strength linearly with strain (from $\sim -20\,\textrm{meV/Fe atom}$ to $\sim -15\,\textrm{meV/Fe atom}$), with a rate of $\Delta J_\textrm{eff}(\varepsilon) = 1.33\,\frac{\textrm{meV/Fe atom}}{\% \textrm{strain}}$. That means that starting from BFO with no strain, collinear order is preferred by the exchange interaction with compressive strain, whereas it is weakened with tensile strain. 
This trend is in agreement with Ref.~\cite{Walden2021}, where the authors also find a linear trend in the largest exchange contribution in our range of strain \cite{Note-on-Walden}.

From Fig.~\ref{Figure: Magnetic trends}~(a) it is evident $J_\textrm{eff}$ deviating from $J_1$. The ratio $\mathcal{F} = J_\textrm{eff}/J_1$ is a good indicator for exchange frustration in magnetic systems \cite{Meyer2020_PhD}, where $\mathcal{F} \sim 1$ describes a slightly exchange frustrated system and ratios for $0< \mathcal{F} << 1$ a strongly frustrated system. For BFO, $\mathcal{F}$ decreases with strain from $\mathcal{F}(\varepsilon = -2\%) = 0.75$ to $\mathcal{F}(\varepsilon = +2\%) = 0.60$. That means, that a slight exchange frustration is induced from compressive to tensile strain.
Note that for $\mathcal{F} \sim 0.2$ \cite{Meyer2019} and $\mathcal{F} \sim 0.3$ \cite{Meyer2020_PhD}, ferromagnetic skyrmions in the absence of magnetic field have been observed and predicted, respectively. These ratios are much smaller than in \BFO and hence, magnetic skyrmions driven by exchange frustration are not expected to occur in BFO. In order to be stabilized, skyrmions and other topological magnetic structures rely on the interplay of magnetic exchange, DMI and anisotropy.

\subsection{Dzyaloshinskii-Moriya Interaction}

We investigate the behavior of the DMI under strain in Fig.~\ref{Figure: Magnetic trends}~(b). As we have shown for bulk BFO \cite{Meyer2023}, we have used the converse spin-current model \cite{Katsura2005,Rahmedov2012} to describe our DFT results for the DMI [Eq.~\eqref{Equation: SC model}]. 
$C_1$ (black points) is decreasing almost linearly from compressive to tensile strain, the variation of $C_2$ (green squares) and $C_3$ (blue triangles) is rather small; however, they have opposite signs at almost the same magnitude for each value of strain. 
Effectively, the DMI (black circles) decreases with strain from $C_\textrm{eff} = 0.33\,\textrm{meV/Fe atom}$ at $\varepsilon = -2\,\%$  to $C_\textrm{eff} = 0.25\,\textrm{meV/Fe atom}$ at $\varepsilon = +2\,\%$. This is a change rate of about $\Delta C_\textrm{eff}(\varepsilon) = -0.02\,\frac{\textrm{meV/Fe atom}}{\% \textrm{strain}}$. That means that compared to $R3c$ BFO, the DMI increases with compression preferring non-collinear order and weakens with tensile strain which lowers the energy for non-collinear order. This is opposite to the trend of the exchange interaction, hence it is important to compare their ratios.

\subsection{Magnetocrystalline Anisotropy}

As third magnetic interaction, we focus on the magnetocrystalline anisotropy [Fig.~\ref{Figure: Magnetic trends}~(c)]. We determine the energy of different crystallographic directions with respect to the hard [111] axis. In Ref.~\onlinecite{Meyer2023}, we have shown that in $R3c$ bulk BFO, the magnetocrystalline anisotropy energy is lowest and degenerate within the full (111) plane. Here, in the absence of strain ($\varepsilon = 0\,\%$), we see the same phenomenon: The calculated directions $[1\overline{1}0]$ (black points), $[11\overline{2}]$ (light green squares), $10\overline{1}$ (purple pentagons) and $[1\overline{2}1]$ (orange upside-down triangles) have the same energy at $-0.07\,\textrm{meV/Fe atom}$ (Note that a tiny deviation is caused by the imposed $Cc$ phase at 0\% of strain). The $[100]$ and the $[001]$ directions (dark green rhombus and lighter green x, respectively) are slightly higher in energy.

Applying strain changes this behavior and the energy degeneracy of the different directions is lifted. For tensile strain $\varepsilon > 0$, the $[1\overline{1}0]$ direction is almost constant in energy while the other directions become unfavorable, most of all the $[11\overline{2}]$ direction.
For compressive strain, the effect is reversed. The $[1\overline{1}0]$ direction gains the least energy, whereas the $[11\overline{2}]$ direction becomes the preferred magnetization direction by the anisotropy. The value increases from $-0.07\,\textrm{meV/Fe atom}$ without strain by $40\,\%$ to $-0.1\,\textrm{meV/Fe atom}$ at $\varepsilon = -2\,\%$. 
We focus on the two directions with lowest energies, $[11\overline{2}]$ and $[1\overline{1}0]$. The behavior of their energies with respect to applied strain differs qualitatively from those of the exchange and DMI of panel (a,b). Here, we can apply a quadratic fit to describe the trend of the anisotropy with strain $K_{[uvw]}(\varepsilon)$. Following the trend for the two interactions, for larger compressive strains, the $[1\overline{1}0]$ direction will be lower in energy than the $[11\overline{2}]$ direction - in accordance with DFT calculations of Ref.~\cite{Chen2018} where the authors also find an easy (111) plane for 0~\% of strain and the easy axis for large compression. However, in the low strain regime, our results slightly differ.

With respect to the $R3c$ phase of BFO, compression leads to an increased MAE, which favors the collinear order of the G-type AFM state. For tensile strain, the MAE remains similar to that of bulk BFO, so a change of the magnetic ground state is not expected from the MAE. In the following, we will discuss the magnetic ground states we derive from our calculations, compare them to current experimental results and show which regime of strain could stabilize potential topological magnetic structures.

\section{Discussion}\label{Section: Discussion}

For over a decade, the different magnetic textures in strained \BFO have been intensively investigated in experiments. However, while several studies show a destruction of the cycloidal order with large compressive strain of BFO on SrTiO\textsubscript{3} substrates (estimated strain of $-1.5\,\% < \varepsilon < -0.5\,\%$ depending on the thickness of BFO) \cite{Bai2005,Bea2007,Zhao2006,Infante2010,Ratcliff2011,Sando2013,Sando2014}, others report the persistence of cycloidal order in the same system \cite{Ke2010,Haykal2020}. For tensile strain, the literature is not equivalently vast, but the destruction of the cycloid with large tensile strain has also been observed \cite{Sando2013,Haykal2020}. There, the amount of applied strain to stabilize collinear order differs by $\sim 1\,\%$. These results show that the magnetic ground state in strained BFO is very sensitive to the thickness and growth conditions. Different cycloids, such as the type-I and type-II cycloids in $[1\overline{1}0]$ and $[11\overline{2}]$ directions, respectively as well as the collinear G-type AFM states are extremely close in energy. Hence, small distortions have a large impact on the magnetic ground state. As we will show in the following, this behaviour is in agreement with our calculations because different magnetic states are within energy ranges down to 10$\mu$eV/atom.

\begin{figure}
\includegraphics[scale=1]{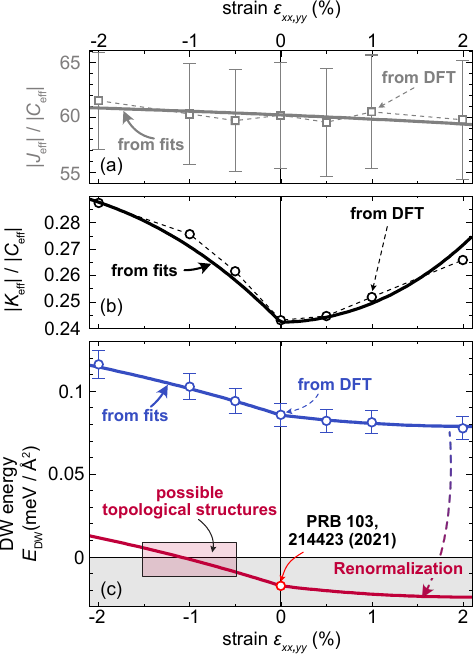}
\caption{Interplay between the magnetic interactions for different values of strain. (a) Ratio of magnetic exchange $J_\textrm{eff}$ with respect to the Dzyaloshinskii-Moriya interaction $C_\textrm{eff}$. Note that the large error estimate stems solely from the small error $\Delta J_\textrm{eff}$. The squares show the values, determined from DFT whereas the solid line shows the ratio \nicefrac{$J_\textrm{eff}(\varepsilon)$}{$C_\textrm{eff}(\varepsilon)$} from the fits of panel (a) and (b) of Fig.~\ref{Figure: Magnetic trends}. (b) Ratio between anisotropy energy $K_\textrm{eff}$ and $C_\textrm{eff}$. The circles show the values, determined from DFT whereas the solid line shows the ratio \nicefrac{$K_\textrm{eff}(\varepsilon)$}{$C_\textrm{eff}(\varepsilon)$} from the fits of panel (c) and (b) of Fig.~\ref{Figure: Magnetic trends}, where $K_\textrm{eff}$ denotes the direction of lowest energy. (c) Domain wall energy $E_\textrm{DW}$ [Eq.~\eqref{Equation: DW energy}] as a function of applied strain. The blue points are derived from the DFT values and the blue solid line shows the trend from the fits of $J_\textrm{eff}(\varepsilon)$, $C_\textrm{eff}(\varepsilon)$ and $K_\textrm{eff}(\varepsilon)$. The red point shows the determined value for the $R3c$ phase of BFO according to Ref.~\cite{Xu2021} where the red line denotes the trend which is renormalized to that point.}
\label{Figure: Magnetic ratios}
\end{figure}

In general, the magnetic ground state is driven by the interplay between the magnetic exchange $J$, the DMI $C$ and the anisotropy $K$. For example, when $K$ is large as compared to $J$ and $C$ a FM or AFM ground state is favored depending on the sign of $J$. When $C$ is large, a non-collinear ground state is favored since the DM favors 90$^{\circ}$ spin spirals. The ratio between the different contributions is therefore important. We compare these interactions in Figure \ref{Figure: Magnetic ratios}. In panel (a), the ratio of $|J_\textrm{eff}|/|C_\textrm{eff}|$ is shown in grey (squares show the ratios of DFT determined magnetic interactions - the bold line is the ratio of the fitted interactions with strain). Increasing the ratios means that collinear order is enhanced, decreasing ratios destabilize collinear magnetic ground states.

Applying compressive strain, the exchange interaction slightly increases with respect to the DMI - collinear G-type AFM order is more favorable than in the absence of strain. With tensile strain, the ratio lowers - hence, non-collinear order gains energy.
The anisotropy [panel (b)] increases for compressive strain from about 24~\% to almost 29~\% of the DMI due to the increasing preference of the $[11\overline{2}]$ direction shown in Fig.~\ref{Figure: Magnetic trends}~(c). This enhances the stability of collinear order with a quantization axis along $[11\overline{2}]$.
With tensile strain, the preferred magnetization direction changes to $[1\overline{1}0]$ with a constant value of the anisotropy. As the DMI decreases, the ratio $\nicefrac{K}{C}$ increases and will also stabilize collinear magnetic order. Note that the increase of anisotropy with respect to DMI starting from  0~\% of strain differs for compressive and tensile strain - e.g. a ratio of 0.26 is reached at $\sim +1.5\,\%$ within the tensile regime, but at around $\sim -0.75\,\%$ in the compressive regime.
Due to both increasing \nicefrac{$J$}{$C$} and \nicefrac{$K$}{$C$} with compression, we expect collinear order to be favored even in low strain regimes compared to the $R3c$ bulk BFO phase. For tensile strain, the conclusion is not that unambiguous: while the exchange loses energy with respect to the DMI, the anisotropy gains energy. The magnetic behavior crucially depends on both trends. In Fig.~\ref{Figure: Magnetic trends}, the effective exchange and DMI depend linearly on the nominal strain, but the anisotropy can be described with a quadratic behavior.

To quantify where isolated chiral magnetic textures may be stable, we have used the domain wall energy, defined as \cite{Craik2003,Lacheisserie2005}
\begin{equation}
    E_\textrm{DW}(\varepsilon) = \frac{4}{a^2(\varepsilon)}\sqrt{J_\textrm{eff}(\varepsilon)\cdot K_\textrm{eff}(\varepsilon)} - \frac{2\pi}{a^2(\varepsilon)} C_\textrm{eff}(\varepsilon), \label{Equation: DW energy}
\end{equation}
with $a(\varepsilon)$ being the pseudo-cubic lattice constant for each value of strain, taken from the diagonal [111] direction to address for the changes in the $c$ axis due to relaxation. We also renormalized $a$ into the $[1\overline{1}0]$ direction to account for the known type-I cycloid.
As compared to other quantities such as the isolated skyrmion energies or vortex energies, the DW energy is independent of the size of the magnetic textures and therefore does not require the expensive spin dynamics energy minimization in a magnetic supercell of at least 100$\times$100 nm at least \footnote{Note that compared to this manuscript, in \cite{Li2023}, the authors assume a [111] directed anisotropy which exceeds the bulk value by a factor of 3 to stabilize magnetic skyrmions in BFO. Here, we have calculated the anisotropy based on DFT and only find an easy axis within (111) plane.}. The DW energy is directly quantifying the energy cost or gain of non collinear textures based on the effective exchange, the anisotropy and the DMI. Furthermore, a domain wall can be approximated as half of a cycloid, where for positive values of energy, collinear magnetic order is preferred and for negative values of $E_\textrm{DW}$, a spin spiral ground state is expected \footnote{Note that contrary to the spin spiral pitch $\lambda$, the DW width $\Delta \propto \sqrt{J/K}$ is independent of the DMI, hence for spin spirals with a certain pitch length the quantitative description differs as well as its energy. However, due to the simplicity of Eq.~(\ref{Equation: DW energy}), the trends for the magnetic ground state energies can be recognized directly and give qualitatively the same results as spin spiral energies.}.

In Fig.~\ref{Figure: Magnetic ratios}~(c), the DW energy with respect to the applied strain is shown. For all values of strain, the DWs are higher in energy than the collinear state (blue points with error bars); however, the energies are extremely small (below 0.1~meV). As described above, compressing BFO prefers collinear order, i.e. an increased DW energy. For tensile strain, the energy differences are almost zero, meaning that we predict a similar ground state compared to the magnetic state at 0~\% strain. The trend coincides with experimental observations: For compressive strain, the collinear G-type AFM state is preferred, driven by the large increase of anisotropy in the $[11\overline{2}]$ direction and the increase in exchange with respect to the DMI. This result is robust against small errors within our calculated interactions. For tensile strain, our results show the same difficulties as for experimental observations: The energies of the states are so close that slightly differing strains, the magnetic ground state can differ (cf. Refs.~\cite{Sando2013} and \cite{Haykal2020}). This is indicated by the error bars from our calculations.

As a reference point in red, we plot the estimated DW energy from Ref.~\cite{Xu2021}, where we investigated the $R3c$ phase of BFO and slightly adapted the DFT parameters to obtain a spin cycloid with a pitch of about $\sim 62\,\textrm{nm}$. With these parameters (corresponding to $C_\textrm{eff} = 0.66\,\textrm{meV/Fe atom}$, $K_\textrm{eff} = 0.069\,\textrm{meV/Fe atom}$ and $J_\textrm{eff} = -11.78\,\textrm{meV/Fe atom}$), the DW energy is slightly negative ($-0.017\,\nicefrac{\textrm{meV}}{\AA}$). Following our estimated trend for the magnetic interactions by renormalizing the DW energy from fits (red curve), we can expect a collinear G-type AFM from $\sim -1.5\,\%$ of compressive strain and a cycloidal ground state for low compressive strains. In between these regimes, at compressive strain $-1.5\,\% < \varepsilon_{xx,yy} < -0.5\,\%$, we expect the transition between non-collinear and collinear order for BFO. This is the region where topologically stable magnetic structures, such as isolated skyrmions or merons and anti-merons could be stabilized in BFO. An investigation of those state is beyond the scope of this paper and part of a different study. 

\section{Conclusion} \label{Section: Conclusion}

We have investigated the evolution of magnetic interactions in \BFO under bi-axial strain between $-2\,\% < \varepsilon_{xx,yy} < +2\,\% $ using first-principles calculations. Going from compressive to tensile strain, both exchange and Dzyaloshinskii-Moriya interaction decrease linearly whereas the uniaxial anisotropy prefers a $[11\overline{2}]$ magnetization direction for the calculated range of compressive strain and a $[1\overline{1}0]$ axis for tensile strain. Comparing the ratios of the three interactions, collinear G-type antiferromagnetic order in \BFO is strengthened for compressive strain due to a large increase of anisotropy by about 40~\% in accordance with experimental observations. For tensile strain our results show a similar magnetic ground state as for bulk BFO, where for a slight change in the ratios, also the collinear ground state can be favored.
At the transition between the cycloid and the collinear AFM state for medium compressive strain more complex magnetic structures such as skyrmions or merons might occur which is why we believe this regime to be interesting for further investigation.

\begin{acknowledgments}
B.D., S.M. thank Prof. Philippe Ghosez, Louis Bastogne, Dr. Subhadeep Bandyopadhyay, Dr. He Xu and Prof. Matthieu J. Verstraete for helpful discussions. This work is supported by the National Natural Science Foundation of China under Grant No. 12074277, the startup fund from Soochow University and the support from Priority Academic Program Development (PAPD) of Jiangsu Higher Education Institutions. S.M., B.D., and L.B. acknowledge the DARPA Grant No. HR0011727183-D18AP00010 (TEE Program) and the European Union’s Horizon 2020 research and innovation programme under Grant Agreements No. 964931 (TSAR). L.B. also thanks ARO Grant No. W911NF-21-1-0113 and ARO Grant No. W911NF-21-2-0162 (ETHOS). Computing time was provided by ARCHER and ARCHER2 based in the United Kingdom at National Supercomputing Service with support from the PRACE aisbl, from the Consortium d’\'Equipements de Calcul Intensif (FRS-FNRS Belgium GA 2.5020.11) and the LUMI CECI/Belgium for awarding this project access to the LUMI supercomputer, owned by the EuroHPC Joint Undertaking, hosted by CSC (Finland) and the LUMI consortium through LUMI CECI/Belgium, ULiege-NANOMAT-SKYRM-1. Sebastian Meyer is a Postdoctoral Researcher [CR] of the Fonds de la Recherche Scientifique – FNRS. Bertrand Dup{\'e} is a Research Associate [CQ]  of the Fonds de la Recherche Scientifique – FNRS.
\end{acknowledgments}

\end{document}